\documentclass[draft,grl]{agutex}
\usepackage{lineno}
\usepackage{graphicx,color,ulem}
\setkeys{Gin}{draft=false}

\authorrunninghead{Y\.I\u G\.IT ET AL.}
\titlerunninghead{GRAVITY WAVES AND CO$_2$ CLOUDS}

\authoraddr{Corresponding author: E. Yi\u git,
Space Weather Group, School of Physics Astronomy and Computational Sciences,
MSN:3F3, Fairfax, VA 22030, USA. (eyigit@gmu.edu)}

\begin{document}
\title{Gravity waves and high-altitude CO$_2$ ice cloud
formation in the Martian atmosphere}
\authors{Erdal Yi\u git, \altaffilmark{1,2} 
Alexander S. Medvedev,\altaffilmark{2,3} 
Paul Hartogh\altaffilmark{2} } 
\altaffiltext{1}{School of Physics, Astronomy, and Computational Sciences, 
George Mason University, Fairfax, Virginia, USA.}  
\altaffiltext{2}{Max Planck Institute for Solar System Research, 
G\"ottingen, Germany.}
\altaffiltext{3}{Institute of Astrophysics, Georg-August University,
G\"ottingen, Germany.}

\begin{abstract}

We present the first general circulation model simulations that quantify 
and reproduce patches of extremely cold air required for CO$_2$ condensation 
and cloud formation in the Martian mesosphere. They are created by 
subgrid-scale gravity waves (GWs) accounted for in the model with the 
interactively implemented spectral parameterization. Distributions of 
GW-induced temperature fluctuations and occurrences of supersaturation 
conditions are in a good agreement with observations of high-altitude CO$_2$ 
ice clouds. Our study confirms the key role of GWs in facilitating CO$_2$
cloud formation, discusses their tidal modulation, and predicts clouds at
altitudes higher than have been observed to date.

\end{abstract}

\begin{article}
\section{Introduction}
\label{sec:intro}

Carbon dioxide ice clouds have been numerously and unambiguously detected in the
middle atmosphere (50--100 km) of Mars \citep[e.g.,][]{Schofield_etal97,
  ClancySandor98, Clancy_etal07,Montmessin_etal07, McConnochie_etal10,
  Vincendon_etal11}. The most recent and detailed references to observations and
climatology of these clouds are given in the works by \citet[Figures 6 to 9 for
distribution over latitude, longitude, $L_s$ and altitude]{SeftonNash_etal13}
and \citet[Table 1 for general characteristics of detected mesosphereic clouds,
and references therein] {Listowski_etal14}.  Observations show that, in average,
temperatures in the Martian mesosphere are not cold enough to sustain CO$_2$
condensation. \citet{ClancySandor98} speculated that CO$_2$ ice forms in patches
of cold air created occasionally by temperature fluctuations associated with
tides and gravity waves (GWs).  This hypothesis has been further supported by
numerical simulations of \citet{Spiga_etal12}. Using a mesoscale (GW-resolving)
Martian meteorological model, they demonstrated that GWs generated in the
troposphere have amplitudes in the mesosphere sufficient to form pockets of cold
air with temperatures below the CO$_2$ condensation threshold. On the other
hand, Martian general circulation models (MGCMs) turned out to be less
successful in reproducing cold temperatures required for the formation of
mesospheric CO$_2$ ice clouds.  Thus, \citet{GonzalezGalindo_etal11} reported
that mesopause temperatures simulated with the Laboratoire de M\'et\'eorologie
Dynamique (LMD) MGCM tended to be low near the equator at $L_s<150^\circ$ and in
middle latitudes afterwards, where high-altitude clouds are usually observed,
albeit always remained few Kelvin degrees above the condensation
point. Similarly, temperatures simulated with the Max Planck Institute (MPI)
MGCM never dropped below that of CO$_2$ condensation \citep[e.g.,][Figure
1]{Medvedev_etal13}.

A possible reason for MGCM failures to reproduce CO$_2$ ice cloud formation in
the mesosphere is that mesoscale GWs (with horizontal scales from tens to few
hundred km) are mostly unresolved by such models. Since dynamical and thermal
effects of GWs in the middle and upper atmosphere are significant and cannot be
neglected, they have to be parameterized in GCMs \citep[see the review of][on
the recent progress with relevant studies in the Earth
atmosphere]{YigitMedvedev15}. On Mars, GW sources are stronger than on Earth,
and significant wave-induced dynamical \citep{Medvedev_etal11a} and thermal
\citep{MedvedevYigit12} effects are being increasingly revealed. In particular,
the simulations of \citet{MedvedevYigit12} with the MPI--MGCM have demonstrated
for the first time that accounting for the GW-induced cooling rates helped to
reproduce the cold mesospheric temperatures retrieved from Mars Odyssey
aerobraking measurements.

In addition to the ability of parameterizations to account for
acceleration/deceleration of the mean flow and mean heating/rates due to
subgrid-scale GWs, they provide amplitudes of GW-induced temperature
fluctuations. In this paper, we extend the study of \citet{Spiga_etal12} from
demonstrating a potential role of GWs in creating pockets of sub-condensation
temperatures to explicitly quantifying and capturing these processes in Martian
GCMs, which open new perspectives of modeling CO$_2$ ice cloud formation. For
that, we employ the MPI--MGCM with the implemented subgrid-scale GW
parameterization of \citet{Yigit_etal08}. The model and GW scheme are described
in Section~\ref{sec:tools}. The simulated global distributions of GW activity
and pockets of air with subcondensation temperatures are presented in
Section~\ref{sec:glob}. Their geographical distributions, and local and
universal time variations are discussed in Sections~\ref{sec:geo} and
\ref{sec:LT-UT}, correspondingly.

\section{Martian General Circulation Model and Gravity Wave Scheme}
\label{sec:tools}

The Max Planck Institute Martian General Circulation Model (MPI--MGCM) is a 
first principle three-dimensional spectral model extending from the surface 
to the lower thermosphere ($3.6\times 10^{-6}$ hPa, $\sim$150-160 km).  
The current simulations have been performed with 67 hybrid levels, and T21 
horizontal resolution (64 and 32 grid points in longitude and latitude,
correspondingly). The model has a complete set of physical parameterizations 
suitable for Mars, which have been described in detail in the works by
\citet{Hartogh_etal05, Hartogh_etal07, MedvedevHartogh07}, and
\citet{Medvedev_etal11b}. In the mesosphere and thermosphere, in the region of
particular interest for CO$_2$ cloud formation, heating/cooling effects due to
radiative transfer in gaseous CO$_2$ are taken into account with the exact
non-LTE ALI-ARMS code of \citet{Kutepov_etal98} and \citet{GusevKutepov03}
optimized with respect to a number of vibrational levels.

The extended nonlinear spectral GW parameterization is described in detail in
the work by \citet{Yigit_etal08}, and its implementation and
application in the MPI-MGCM is given in the paper by
\citet{Medvedev_etal11b}. The scheme is appropriate for planetary whole 
atmosphere models extending into the thermosphere. It has been extensively 
tested and used for studying GW-induced vertical coupling processes in Earth's 
atmosphere \citep{Yigit_etal09,Yigit_etal12b,Yigit_etal14,YigitMedvedev09,
YigitMedvedev10}. The scheme calculates vertical propagation of subgrid-scale
harmonics systematically accounting for the major wave dissipation 
mechanisms in the Martian atmosphere: due to nonlinear breaking/saturation, 
molecular viscosity and thermal conduction. An empirical distribution of GW 
spectrum is specified at the source level in the lower atmosphere, and the 
scheme calculates at consecutive vertical levels the momentum and temperature 
tendencies imposed by GWs on the larger-scale (resolved) atmospheric flow.  
In this study, we use the same GW source specification as in the previous works 
\citep{MedvedevYigit12, Medvedev_etal13}. Namely, the horizontally
uniform and temporally constant amplitude of GW velocity fluctuations ($\sim$1 
m~s$^{-1}$) was prescribed at the source level ($p=260$ Pa), harmonics were
launched along and against the direction of the local wind, and a Gaussian 
spectrum over horizontal phase speeds was assumed.
  
The model was run for 40 sols, approximately corresponding to $L_s=0-20^\circ$,
that is, for a Northern Hemisphere spring equinox. The low dust optical depth
($\tau=0.2$ in visible) appropriate for this season, and the low solar flux of 
$F_{10.7}= 80\times10^{-22}$ W m$^{-2}$ Hz$^{-1}$ have been kept constant 
throughout the simulations. Model mean fields to be shown have been calculated 
as 60-sol averaged, and the universal time plots are based on hourly outputs.

\section{Gravity Wave Activity and Subsaturation Temperatures}
\label{sec:glob}

The GW parameterization calculates vertical profiles $|u^\prime_j(z)|$ of
amplitudes of horizontal velocity fluctuations for each wave harmonic $j$ of 
the GW spectrum. The contributions of all harmonics yield the root-mean-square
(RMS) $\sigma_u(z)=\sqrt{\sum_j |u^{\prime}_j(z)|^2}\equiv |u^\prime|$, which
describes the net GW-induced variance of horizontal velocity. Because the
kinetic ($E_k$) and mechanical potential ($E_p$) energies are equal for GW 
harmonics under consideration \citep{GellerGong10},
  \begin{equation}
    E_k=\frac{1}{2}|u^{\prime}|^2=
    \frac{1}{2}\biggl( \frac{g}{N} \biggr)^2 \frac{|T^{\prime}|^2}{T^2} = E_p,
  \end{equation}
where $g$ is the acceleration of gravity, $N$ is the Brunt-V\"ais\"al\"a
frequency, and $T$ is the mean (resolved by the GCM) neutral temperature, the 
total wave-induced RMS temperature perturbations $|T^\prime|$ can be computed 
from $|u^\prime|$. Note that $|T^\prime|$ provides no information about phases
of contributing harmonics. Unlike with wave-resolving simulations, 
$|T^\prime(z)|$ does not describe instantaneous wave fields, but only 
characterizes the magnitude of the parameterized subgrid-scale GWs. 
Individual wave-induced temperature fluctuations in each point are, thus, 
bound by $|T^\prime|$: $-|T^\prime| \le T^\prime \le |T^\prime|$.

\begin{figure}
  \centering
  \includegraphics[clip,width=1\columnwidth]{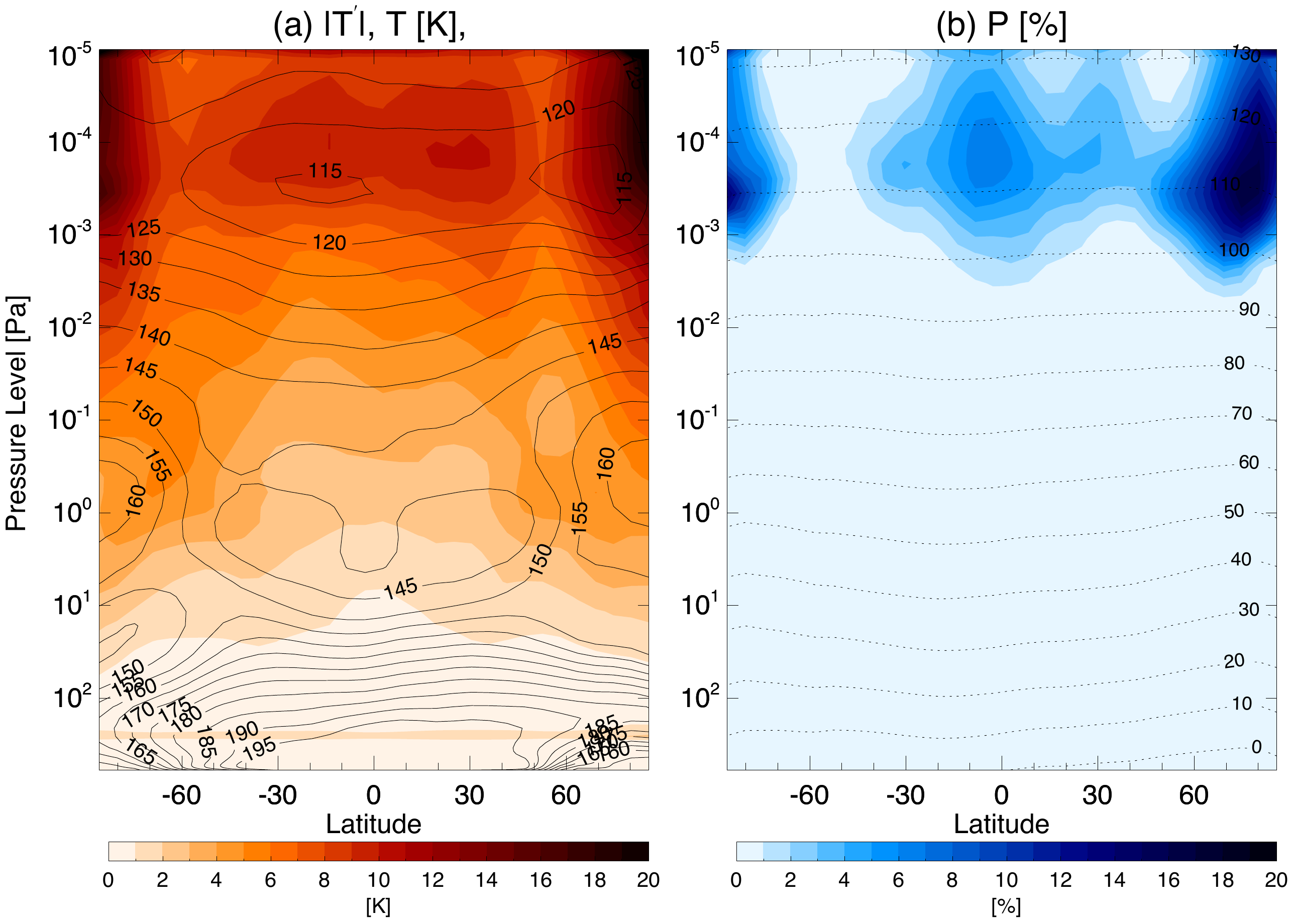}
  \vspace{-1cm}
  \caption{Mean (40-sol) (a) zonal mean neutral temperature $T$ (contour lines)
    and gravity wave-induced temperature perturbation $|T^\prime|$ (color
    shaded) in Kelvin degrees; (b) probability $P$ of CO$_2$ ice cloud formation
    in percentage (blue shaded), and the geopotential height of the corresponding
    pressure levels in km (contours).}
  \label{fig:fig1}
\end{figure}

The simulated temperature $T$ (black contours) and GW activity $|T^\prime|$
(color shades) are plotted in Figure~\ref{fig:fig1}a in the form of zonally and
temporally averaged pressure-latitude cross-sections. The GW activity varies
appreciably both with height and latitude. Temperature fluctuations increase
with altitude from about 2 K near the source in the troposphere (at 260 Pa),
peak near the mesopause ($\sim$105 km) with $|T^\prime|\sim 12$ K around
low-latitudes, and reach 20 K in the polar regions in both
hemispheres.  Enhanced GW activity in the mesopause regions occurs as a result
of exponential amplitude growth due the density decay, and the attenuation of
upward propagating lower atmospheric GWs by nonlinear diffusion and increasing
molecular viscosity. Selective filtering of GW harmonics by the winds is also 
at play, and is mainly responsible for the latitudinal distribution of
$|T^\prime|$.

CO$_2$ condensation requires a temperature drop below the saturation level,
which is calculated in our model with the Clausius-Clayperon relation as the
function of pressure. Our GCM does not include a detailed cloud microphysics. 
Instead, the process of ice particle nucleation is accounted for by a 
significant degree of supersaturation: CO$_2$ ice is considered to be formed 
at pressures 1.35 times larger than that from the Clausius-Clayperon relation, 
as suggested by \citet{Glandorf_etal02} and utilized by \citet{Kuroda_etal13}. 
As already mentioned, temperature $T$ in our simulations never fell below thus 
defined saturation level $T_s$, which can also be seen in 
Figure~\ref{fig:fig1}a. However, taking excursions around the resolved
temperature induced by parameterized small-scale GWs into account can yield 
the necessary CO$_2$ ice condensation conditions. Having no way of knowing
individual wave profiles, one nevertheless can argue that condensation may 
occur only when $T-|T^\prime| \le T_s$. We count these events at every time step 
over the course of model integration, and their number (divided by the total 
number of time steps) defines the probability $P$ of CO$_2$ ice cloud 
formation over the time period. Alternatively, the value of $P$ shows the 
percentage of time when super-saturation conditions exist. 

Figure~\ref{fig:fig1}b presents thus calculated $P$ in percentage (shaded), and
the geopotential height in km is overplotted with contours for more convenient
visualization of geographical altitudes. It is seen that chances of encountering
CO$_2$ super-saturation conditions increase rapidly with height from the upper
mesosphere upward. Visual inspection of panels a and b reveals that there is a
remarkable correlation between the GW-induced temperature perturbations and the
chances of CO$_2$ clouds to form. The largest probabilities are around the
mesopause ($P$ up to 10\% at low-latitudes and exceed 20\% at
high-latitudes). More pockets of cold air with subcondensation temperatures in
these regions are due to the combined effect of globally lowest temperatures and
largest GW perturbations. In the upper mesosphere $P$ is up to 5\% around the
equator. It approximately matches the region
where the CO$_2$ clouds have been observed over this season \citep[e.g.,][and
references therein]{SeftonNash_etal13}, while capturing the fact that the cloud
events are relatively rare. 
If subgrid-scale GW effects are not included in simulations, the 
supersaturation of CO$_2$ does not occur at all, that is, $P= 0$ globally 
(not shown).

These results indicate that (1) if subgrid-scale GW effects are included, then
the supersaturation condition is fulfilled globally in up to $>20\%$ of the time
in the mesosphere and mesopause region over a large range of altitudes and
latitudes; (2) regions of most probable cloud formation coincide with the places
of enhanced GW activity.  A note of caution should be added here. Due to the
parameterized nature of GWs in our simulations, the occurrences of clouds are
not deterministic.  Probability $P$ quantifies regions where CO$_2$ condensation
is most likely, but individual temperature profiles may not yield temperatures
below $T_s$ due to particular phases of the harmonics and their interference.

\section{Mean Geographical Distribution}
\label{sec:geo}

\begin{figure}
  \centering
  \includegraphics[width=1\columnwidth]{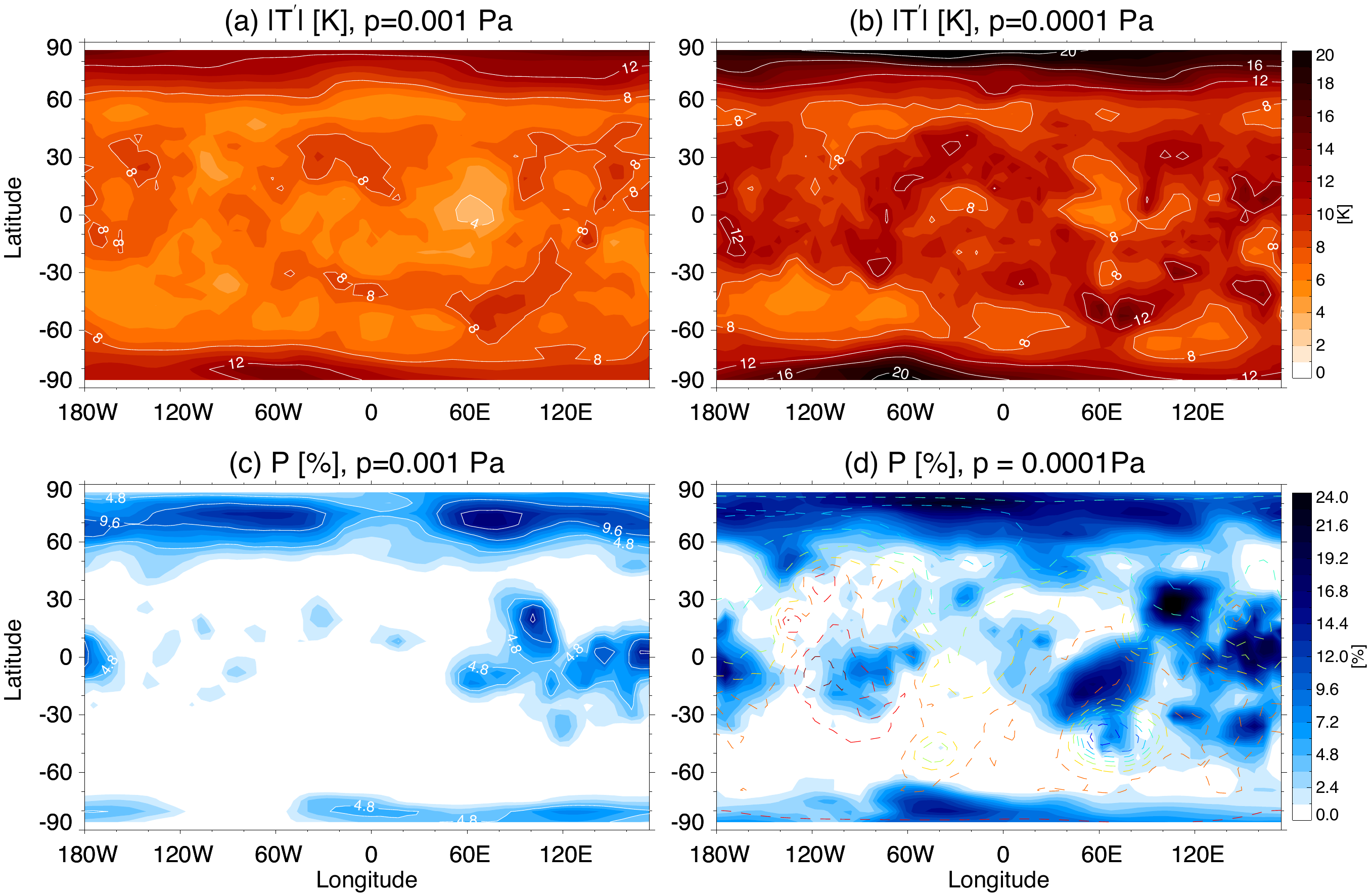}
  \vspace{-1cm}
  \caption{Geographical distribution of the mean gravity wave-induced
    temperature perturbations $|T^\prime|$ in K (upper panels) and the 
    probability of CO$_2$ ice cloud formation $P$ in percentage (lower 
    panels) at $10^{-3}$ (left) and $10^{-4}$ Pa (right). Dashed lines 
    in the lower panels show the main topographical features from low 
    (blue) to high levels (red).}
  \label{fig:fig2}
\end{figure}

We next investigate the geographical details of the modeled link between the
GW-induced temperature fluctuations $|T^\prime|$ and the probability of CO$_2$
ice cloud formation $P$. Figure~\ref{fig:fig2} presents the mean
latitude-longitude cross-sections of $|T^\prime|$ (upper panels) and $P$ (lower
panels) at two representative pressure levels in the upper mesosphere and around
the mesopause, $p=10^{-3}$ Pa (left) and 10$^{-4}$ Pa (right),
respectively. Major topographic features are indicated in the lower panels with
blue-to-red (that is, low-to-high levels) contour scales.

There is a significant amount of spatial variability in the distribution of
$|T^\prime|$ at both altitudes, in particular, around the mesopause. In the
mesosphere, GW-induced temperature variations can reach 4--12 K, while higher up
around the mesopause, $|T^\prime|$ is up to $>20$ K. Distributions of cloud
formation probability $P$ show also a marked spatial variability with larger
values concentrated mainly around the high-latitudes and the equator in the
mesosphere with peak values of $P\sim2-3\%$ around $60^\circ$W--$120^\circ$W,
and $P\sim14 \%$ around 60$^\circ$E--160$^\circ$E. These regions partially
coincide with those where clouds have been observed more often \citep[Figure
2]{Spiga_etal12}.  There is an appreciable probability of cloud formation around
the mesopause region. Very large values of $P$ are seen, in particular, at
low-latitudes and near both poles with values exceeding 24\% locally. Overall,
the results indicate that there exists a very good geographical correlation
between the GW activity, and the chances of mesospheric CO$_2$ ice clouds to
form, especially, near the mesopause. Locations of such spots of high wave
activity (and cloud formation as well) are not clearly linked down to surface
features. This is because the GW parameterization accounts for ``non-orographic"
waves, which propagate to the mesosphere and lower thermosphere more favorably
than their topographically forced counterparts with very slow observed phase
speeds. The regions of enhanced $|T^\prime|$ are apparently the results of
modulation of GW propagation by larger-scale meteorological features in the
underlying layers, by quasi-stationary waves, for instance.

\section{Local Time and Universal Time Variations}
\label{sec:LT-UT}

\begin{figure}
  \centering
  \includegraphics[width=1\columnwidth]{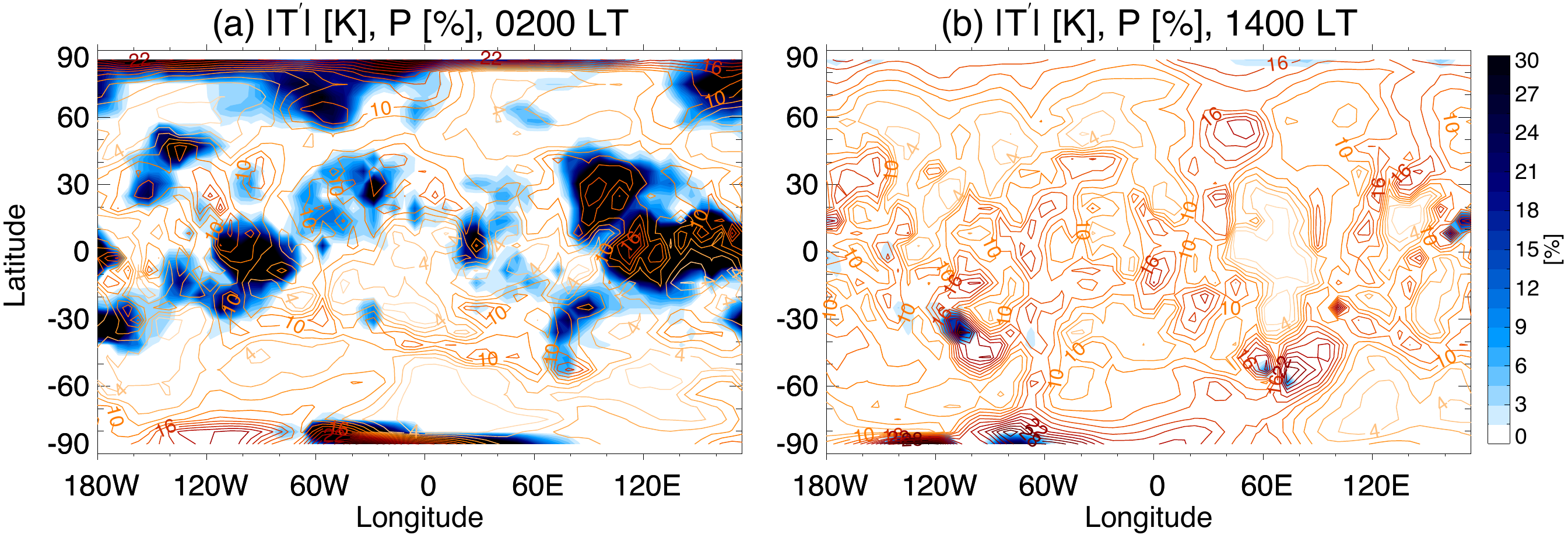}
  \vspace{-0.4cm}
  \caption{Latitude-longitude cross-sections of the 
    simulated GW-induced temperature
    fluctuations (contours), and the probability $P$ of CO$_2$
    supersaturation (blue shaded) around the mesopause (p=0.0001 Pa) during a)
    local night (0200 LT), and b) local afternoon (1400 LT). Contour intervals
    are 3 K.}
  \label{fig:fig3}
\end{figure}

The majority of observations of mesospheric CO$_2$ ice clouds have been 
performed from spacecraft inserted on polar orbits (Mars Global Surveyor, Mars 
Express, Mars Odyssey, Mars Reconnaissance Orbiter). This means that the 
measurements were always taken at narrow intervals of local times, mainly on a 
day side of the planet. To investigate how this may bias the cloud statistics, 
we plotted the simulated distributions at certain local day- and night-times, 
as if they were observed from such an orbiter. Figure~\ref{fig:fig3} presents 
the latitude-longitude distributions of the GW-induced temperature 
perturbations $|T^\prime|$ (contours) and probability of CO$_2$ supersaturation 
$P$ (blue shadings) at 0200 LT (left) and 1400 LT (right) in the mesopause 
region at $p=0.0001$ Pa, where the coldest mean temperatures and largest 
GW-induced temperature fluctuations coexist.

It is immediately seen that more clouds have chances to form at night-times.
Because most of cloud observations took place on the day-side of the orbits, 
our results may serve as an indication that the amount of such clouds on Mars 
is underestimated. Magnitudes of GW activity $|T^\prime|$ are approximately 
equal throughout all local times, however mean temperatures $T$ vary 
significantly. With colder night-side $T$, the supersaturation condition 
$T-|T^\prime| \le T_s$ occurs more often, while local enhancements of 
$|T^\prime|$ and $P$ still substantially correlate.

\begin{figure}
  \centering
  \includegraphics[width=1.0\columnwidth]{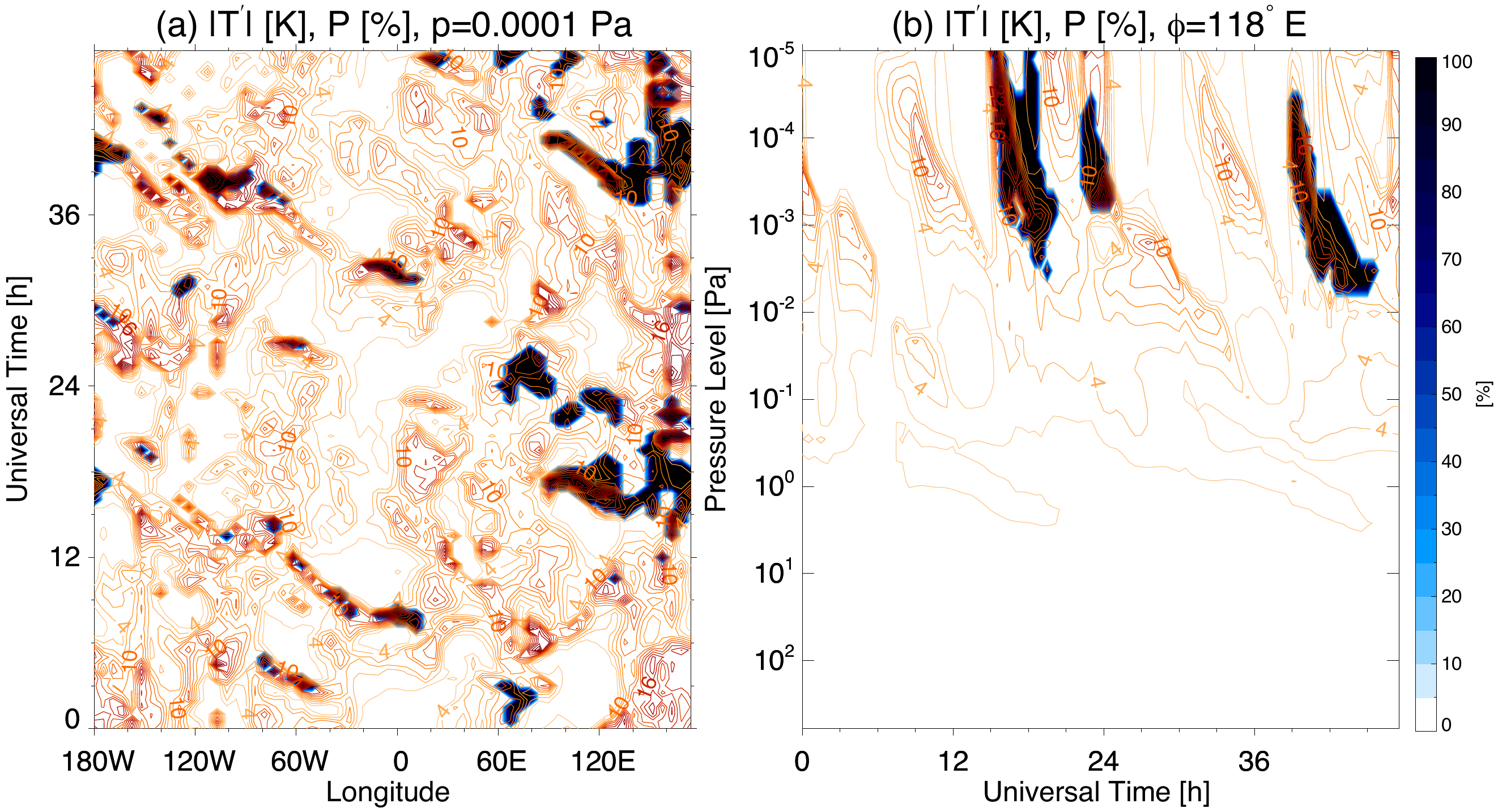}
  \vspace{1cm}
  \caption{Temporal variations (over 48-hour period) of the modeled GW-induced
    temperature perturbations (contours), and the
    probability of CO$_2$ ice cloud formation (blue shading)
    close to the equator ($\theta=2.7^\circ$N): (a) Universal time 
   (UT)-longitude distributions; (b) pressure level (in Pa) - UT 
   distributions at $118^\circ$E.}
  \label{fig:fig4}
\end{figure}

Further insight into how subgrid-scale GWs mediate CO$_2$ ice cloud formation
can be gained from Figure~\ref{fig:fig4}. It shows universal time variations of
the simulated wave activity $|T^\prime|$ and probability of condensation
$P$. The left panel presents the Hovm\"oller diagram (longitude vs time) for
these quantities where they are the largest in our simulations: over the equator
at around the mesopause level ($p=0.0001$ Pa). It demonstrates that cloud
occurrences are extremely localized in space and time, and may appear as almost
random. However, westward propagating variations of $P$ with clear migrating
diurnal and semidiurnal tide signatures can be seen. Regions of high wave
activity are also modulated by the diurnal and semidiurnal solar tides. Although
there is a strong correlation between $|T^\prime|$ and $P$, not all maxima of GW
activity are accompanied by CO$_2$ condensation conditions. The right panel
displays this with the height-universal time cross-sections. Tides affect GW
propagation mainly by altering zonal winds. Therefore, maxima of GW activity
(contours) are phase-shifted with respect to peaks of tidally modulated
temperature $T$.  Condensation occurs only when and where $T$ and $|T^\prime|$
appropriately match the saturation condition (shaded areas).  Thus, our
simulations provide further illustration for the hypothesis of
\citet{ClancySandor98} regarding the combined role of tides and GWs in
mesospheric CO$_2$ cloud formation.

\section{Conclusions}
\label{sec:conc}

We presented the first simulations with a general circulation model of
distributions of localized patches of air at temperatures below the
CO$_2$ condensation threshold. They have been performed with the Max Planck
Institute Martian General Circulation Model (MGCM) with the implemented
subgrid-scale gravity wave (GW) parameterization of \citet{Yigit_etal08}.
The main conclusions are the following.

\begin{itemize}
\item
GWs facilitate cloud formation by creating pockets of cold air with
temperatures below that required for CO$_2$ condensation. Without GWs,
high-altitude cloud formation is impossible, at least in our simulations.

\item
The simulated occurrences of supersaturated temperatures are in a good
agreement with observations of CO$_2$ clouds in the mesosphere.

\item
The approach with utilizing parameterized GWs can be used for quantifying
the mechanism of mesospheric CO$_2$ cloud formation, and reproducing it
in MGCMs.

\item
Our study predicts more clouds than observed at higher altitudes, and in polar
regions.

\end{itemize}

The last prediction requires a discussion, which we have left to this point.
Why the even-higher-altitude CO$_2$ clouds have not been observed to date?
One possibility is that they cannot form for microphysical reasons due to
an insufficient amount of nucleation particles, etc. This can be explored
by implementing a state-of-the art microphysics scheme, and using the approach
described in this paper. The second possibility is that such clouds are too
thin, sizes of ice particles are too small, and the existing instrumentation
was simply unable to detect them. Finally, our prediction can be altered 
when more observations become available to constrain gravity wave sources in 
the lower atmosphere and atomic oxygen that determines CO$_2$ IR cooling in 
the mesosphere \citep{Medvedev_etal15}. This prediction provides a testbed for
assessing our understanding of the Martian middle atmosphere physics.

\begin{acknowledgments}
  Data supporting the figures are available from EY (eyigit@gmu.edu).  EY was
  partially funded by NASA grant NNX13AO36G.  The work was partially
  supported by German Science Foundation (DFG) grant ME2752/3-1.
\end{acknowledgments}

\end{article}
\newpage
 
\end{document}